\documentstyle[11pt,aaspp4,graphicx]{article}

\lefthead{Bock, Turtle and Green}
\righthead{Vela SNR survey}

\newcommand{\cosec}{\mathop{\rm cosec}\nolimits}
\newcommand{\cd}{\hbox{$\cosec\left|\delta\right|$}}
\newcommand{\jb}{\hbox{Jy~beam$^{-1}$}}
\newcommand{\hi}{H\,\textsc{i}}
\newcommand{\hii}{H\,\textsc{ii}}
\newcommand{\hal}{H$\alpha$}
\newcommand{\h}{\hbox{$^{\rm h}$}}
\newcommand{\m}{\hbox{$^{\rm m}$}}

\begin{document}
\title {A high-resolution radio survey of the Vela supernova remnant}

\centerline{(To appear in \emph{The Astronomical Journal})}

\author{D. C.-J. Bock}
\affil{Radio Astronomy Laboratory, University of California, Berkeley,
  CA 94720; and School of Physics, University of Sydney, NSW 2006,
  Australia; dbock@astro.berkeley.edu}

\and

\author{A. J. Turtle and A. J. Green} 
\affil{School of Physics, University of Sydney, NSW 2006, Australia;
  turtle@physics.usyd.edu.au, agreen@physics.usyd.edu.au}

\begin{abstract}
This paper presents a high-resolution radio continuum
(843~MHz) survey of the Vela supernova remnant.  The contrast
between the structures in the central pulsar-powered nebula of the
remnant and the synchrotron radiation shell allows the remnant to be
identified \emph{morphologically} as a member of the composite
class. The data are the first of a composite remnant at spatial
scales comparable with those available for the Cygnus Loop and the
Crab Nebula, and make possible a comparison of radio, optical and
soft X-ray emission from the resolved shell filaments.  The survey,
made with the Molonglo Observatory Synthesis Telescope, covers an
area of 50 square degrees at a resolution of $43''\times60''$,
while imaging structures on scales up to 30\arcmin.
\end{abstract}


\section{Introduction}

The Vela supernova remnant (G$263.9-3.3$) is one of the closest and
brightest supernova remnants.  Recent measurements support a distance
of about 350~pc (\cite{dub98} and references therein).  Estimates of
its age range from a few thousand years (\cite{sto80}) to 29,000~yr
(Aschenbach, Egger, \& Tr\"umper 1995\nocite{asc95}) with a widely
used value being that given by the characteristic age of its pulsar,
11,400~yr (Reichley, Downs \& Morris 1970\nocite{rei70}).  Its
brightness and large angular size ($\sim8$\arcdeg) have made possible
its study at many wavelengths.  Yet it has been less intensively
studied than other close remnants such as the plerionic Crab Nebula
and the shell-type Cygnus Loop, although it is arguably the closest
and a key member of the third major class of SNRs, the composite
remnants. Among many controversies surrounding its nature has been
that of whether it is a shell or a composite remnant. This paper aims
to show that the Vela SNR definitely can be seen as a composite
remnant on morphological grounds and ought to be considered the
Galactic archetype.

Previous radio studies of the Vela SNR region have mainly used lower
resolution single dish images over a frequency range from 408~MHz to
8.4~GHz (\cite{mil68}; Day, Caswell, \& Cooke 1972\nocite{day72},
\cite{mil80}; \cite{mil95}; \cite{dun96}). The earliest observations
showed the Vela SNR to comprise three main areas of radio emission,
called Vela~X, Y, and Z, within a diameter of 5\arcdeg, corresponding
roughly with the bright, filamentary structure of the nebula Stromlo
16 (\cite{gum55}; \cite{mil68a}).  Until recently, this 5\arcdeg\ 
diameter was thought to indicate the extent of the remnant, with the
nebula Vela~X containing the pulsar (PSR B0833--45), offset to one
side. This pulsar, discovered in 1968, was immediately associated with
the Vela supernova remnant (\cite{lar68}).  Calculations by Bailes et
al.\ (1989\nocite{bai89}) indicate that there is only a 0.1\%
probability that the pulsar and the supernova remnant are in chance
superposition.  Kundt (1988\nocite{kun88}) deduced from the 408~MHz
survey of Haslam et al.\ (1982\nocite{has82}) that the Vela SNR might
be much larger, in fact about 8\arcdeg\ across, with the pulsar
approximately at the center. ROSAT and radio observations
(\cite{asc92}; \cite{dun96}) reinforced this model. The discovery that
the speed and direction of the pulsar's proper motion indicates that
it was born near the center of the 8\arcdeg\ Vela SNR shell
(\cite{asc95}) solves the offset problem. This is one of only a few
reliable SNR/pulsar associations (\cite{kas96}). The most recent
single-dish observations (\cite{mil80}; \cite{mil95}) began to resolve
Vela~X at higher frequencies (5--8.4~GHz) and uncovered strongly
linearly polarized structure.  Higher resolution observations at 327
MHz with the Very Large Array (VLA) by Frail et al.\ 
(1997\nocite{fra97}) showed a bright filament near the center of
Vela~X, near the X-ray feature called a `jet' by Markwardt \&
\"Ogelman (1995\nocite{mar95}).  This feature extends southwards from
the pulsar, which is offset to the north of Vela~X.

The radio spectral index of Vela~X is flatter than that of the rest of
the remnant, leading to the remnant's classification as a composite,
with the Vela~X nebula directly powered by the Vela pulsar
(\cite{dwa91}; \cite{wei88}).  This conclusion has been controversial
(\cite{wei80}; \cite{mil86}; \cite{wei88}). The Vela SNR lies close to
the Galactic Plane, leading to difficulties in estimating the
baselevel both in single-dish and interferometer images.  To provide
the first evidence supporting the classification as a composite on
morphological grounds, radio observations are presented in this paper
at the highest resolution yet used to image the Vela X region.  A
subsequent paper will present a multi-wavelength study of Vela~X and
consider the nature of the central plerion in detail.

These are the first high-resolution radio observations to cover a
large fraction of the entire Vela SNR, and are the first radio
observations of the Vela shell at a resolution compatible with
currently available optical and X-ray images.  Radio observations of
the Vela shell published before the present work have been at
relatively low resolution. For example, the observations of Duncan et
al.\ (1996\nocite{dun96}) have a resolution of 10\arcmin, making them
difficult to correlate with high resolution data in other spectral
regimes. In this paper it is possible for the first time to present a
multi-wavelength study of part of a composite remnant's shell at
scales a small fraction of a parsec.

\section{Observations}
\label{sec:surv_obs}

The Molonglo Observatory Synthesis Telescope (MOST) is an east-west
multi-element interferometer located in New South Wales, Australia
(\cite{rob91}). It consists of two co-linear cylindrical parabolas,
each 11.6~m by 778~m. In a twelve-hour synthesis observation it images
an elliptical field of size $70'\times
70'\cosec\left|\delta\right|$.\footnote{Recent modifications have
  increased the field of view to
  $163'\times163'\cosec\left|\delta\right|$.} Sixty-three of these
synthesis observations covering an area of almost 50 square degrees
comprise this survey. The survey includes the regions of brightest
radio emission from the Vela SNR and covers the majority of the X-ray
remnant as seen by Aschenbach et al.\ (1995\nocite{asc95}).  The area
close to the strong \hii\ region RCW~38 has been avoided, due to
imaging artefacts.  Each observation
was made at a frequency of 843~MHz and a resolution of
$43''\times43''\cosec\left|\delta\right|$, receiving right-handed
circular polarization in a bandwidth of 3~MHz.  The observations were
made over the period 1984 February 3 to 1996 February 3.  Complete
coverage with the elliptical field shape necessitates substantial
overlaps in the survey which have been used to refine the relative
calibration between fields, based on the unresolved sources common to
more than one field.  Twenty-seven of the earlier observations were
made for the First Epoch Molonglo Galactic Plane Survey (\cite{whi89};
Green, Cram, \& Large 1998\nocite{gre98}) and some of these data
appear in the MOST supernova remnant catalogue (\cite{whi96}). The
remaining observations, which maintained the same basic grid
separation of 0\fdg9, commenced on 1992 January 17.

Those observations initially made in the vicinity (within about
1$^\circ$) of the Vela pulsar were severely limited in dynamic range
by the presence of the pulsar in the primary beam.  The pulsar is a
strong, unresolved continuum source and is variable over time scales
of seconds to hours. Its integrated pulsed emission in ungated
observations was 2.2$\pm$0.4~Jy, averaged over the entire pulsar period. A
source will appear in a MOST image with a symmetric point-spread
function only if its intensity remains constant throughout the
observation.  Any time-dependent variation compromises sidelobe
cancellation during synthesis, producing rays within the image
emanating from the source, and confusing nearby faint features.  To
improve the imaging of this region a method of pulsar gating was used
which was originally developed by J.~E. Reynolds (personal
communication, 1996) for an observation in 1986.

To make the observations, a predicted pulse arrival time was used to
generate a gating signal of width 20~ms in each pulsar period ($\sim$
89~ms).\footnote{The predicted pulse arrival time was determined from
  timing measurements of the pulsar kindly provided by R.~N.
  Manchester (ATNF), M.~Bailes (University of Melbourne) and P.~M.
  McCulloch (University of Tasmania).} This was displayed on an
oscilloscope simultaneously with the actual detected total power
signal of the pulsar.  The half-power pulse width of the pulsar signal
was 4.5~ms, larger than the observed half-power width of 2.3~ms at
632~MHz (\cite{mcc78}) due to the dispersion (2~ms) over the MOST's
bandwidth (following Komesaroff, Hamilton, \& Ables
1972\nocite{kom72}) and to the effect of an integrating low pass
filter present in the signal path before the detection point.  The
gating signal was adjusted to suppress data acquisition from 5~ms
before until 15~ms after the peak of the pulse, to allow removal of
almost all of the variable pulsed emission. Four observations were
made, each with the pulsar near the edge of the field to the north,
south, east and west. These observations were incorporated in the
complete mosaic in place of the non-gated data, for this part of the
survey. The observations made with this procedure contain only a
40~mJy (2\%) residual of the pulsed emission, sufficient to preclude
associated imaging artefacts.  However, some artefacts are present in
more remote fields, mainly due to the grating response of the MOST to
the pulsar. Further details of the pulsar gating procedure are given
by Bock (1997\nocite{boc97}).

\section{Imaging}
\label{sec:mosaic_imaging}

Individual images were synthesized from each of the 63 (12 hour) observations using the
back-projection algorithm (\cite{per79}; \cite{cra84}) as implemented in the
standard MOST reduction software.  To provide initial calibration for
the reduction process, several strong unresolved sources were observed briefly
before and after each target observation.  These provide a
preliminary gain, phase and beamshape calibration.
The images were deconvolved using the H\"ogbom CLEAN algorithm. Some
images containing stronger sources were adaptively deconvolved as
described by Cram \& Ye (1995\nocite{cra95}).  This method is similar
to self-calibration, but with a reduced set of free parameters.
The residual images had pixel rms in the range 1--6~m\jb. This range
reflects the variation between fields which were essentially noise-limited
and those which were dynamic-range limited.

Preliminary position and flux calibration used short observations of a number (typically eight) of
unresolved calibration sources before and after each Vela target
field. These sources are a subset of the MOST calibrators from
Campbell-Wilson \& Hunstead (1994\nocite{cam94}). A refined calibration was
achieved using unresolved survey field sources which are located in the
overlapped regions. Measurement of these sources produced small corrections
that result in positions mostly consistent to better than $1''$ in right
ascension and $2''$ in declination and flux densities accurate to
within 5\%.\footnote{The uncertainties in the
  absolute calibration of MOST images are approximately an additional 5\% in
  flux density and 0\farcs5 in RA and Dec.  (Hunstead, 1991 and
  personal communication, 1997).}  The refining corrections could not be
applied to some fields which had too few unresolved sources in common with
nearby fields (for example, those at the edge of the mosaic). The
individual fields were mosaiced into ARC projection (\cite{gre95}), to
facilitate comparison with optical surveys.

The MOST is a redundant array sensitive to the
full range of spatial frequencies within the limits set by its extreme
spacings. The minimum geometric spacing (42.85$\lambda$), implies a
maximum detectable angular scale of about 1.3\arcdeg. However, it has
been found empirically that the actual synthesized beam of the MOST is
best fit with a model including a effective minimum spacing of about
70$\lambda$. This parameter varies between and during observations.
Typically, angular scales less than about 30\arcmin\ are well
imaged. The MOST's synthesized beam can also vary significantly
during an observation due to environmental effects and minor telescope
performance variations with the result that the theoretical beam used
for deconvolution sometimes does not model well the actual beam of an
observation. A combination of these effects produces a negative bowl
artefact around bright extended sources.  In the MOST Vela SNR
survey, this effect causes a background level of about
$-10$~m\jb\ around Vela~X and Puppis~A.

Extended structure at levels as low as 6~m\jb\ is clear in the
images.  Although the rms in individual pixels is around 2~m\jb,
typical extended structure covers many pixels and is reliably detected
at lower levels than would usually be accepted for point sources. The
confirmation of this low level structure in a VLA image of Vela~X made at
1.4~GHz (\cite{boc98a}) validates similar emission
seen at 843~MHz elsewhere in the survey.

The most common artefacts present in the image are grating rings, which are
due to the periodic nature of the MOST. They are sections of ellipses of
dimension $1\fdg15\times1\fdg15$\,\cd\ with a width dependent on the source
producing them and are, in general, of variable amplitude
since they pass through several individual fields where the grating
rings have different relative gains.  The morphology of these artefacts
makes them easily distinguishable from the sky emission.  Much of the
survey is dynamic-range limited; in the less complicated regions the
rms noise is of order 1--2~m\jb.

An additional non-ideality comes from the mosaicing: the image from each of
the 63 individual observations was deconvolved separately, yet
structure outside a given field can contribute to sidelobes in that
field.  This manifests itself as discontinuities in the survey image,
at the edges of component images or where
regions containing artifacts have been masked.

\section{Results}

An image of the complete MOST survey of the Vela SNR is shown in
figure~\ref{fig:survey}.  To assist in identifying the various objects
and emission regions within the survey, a cartoon covering the same
area is presented in figure~\ref{fig:vela_cartoon}.   Key
characteristics of the survey are summarized in Table~\ref{params}.

\begin{table}[t]
\caption{\protect\centering Key parameters of the MOST Vela SNR survey
  \label{params}}
\vspace*{0.5mm}
\centering
\begin{tabular}{ll}
\tableline
\tableline
Parameter  &  Value \\
\tableline
frequency    &   843~MHz\\
bandwidth    &     3~MHz\\
total area  & 50~deg.sq.\\
individual field size ($\alpha\times\delta$)   &   $70'\times 70'~$\cd\\
resolution ($\alpha\times\delta$)  &  $43''\times 43''~$\cd\\
max. imaged angular scale&$\sim 30'$\\
typical rms noise & 1--2~m\jb\\
received polarization &  right circular (IEEE)\\
\tableline
\end{tabular}
\end{table}

Morphologically, there are several distinct regions apparent in the
image. Near the center is the bright nebula known as Vela~X, which is
thought to be powered by the Vela pulsar (PSR~B0833$-$45: Large,
Vaughan, \& Mills 1968\nocite{lar68}).  The nebula is composed of a
network of complex filaments. Significant extended structure is also
present but not detected by the MOST. This region is seen more clearly
in subsequent images.

In the north and east of the image there are several filaments from
the shell of the Vela supernova remnant and at least one unrelated
\hii\ region, RCW~32 (Rodgers, Campbell, \& Whiteoak
1960\nocite{rod60}).  There are also partial elliptical artefacts due
to strong \hii\ regions outside the survey area.  Broadly speaking, we
can categorize the extended structure in this area on morphological
grounds. To the north-east of the Galactic Plane, much of the
structure is diffuse and randomly oriented and may be Galactic
emission unrelated to the Vela SNR.  Most of the extended emission
between the Galactic Plane and the Vela~X nebula \emph{is} due to the
Vela supernova event. These filamentary features have some
correspondence with optical filaments and X-ray emission in the area
(\S~\ref{sec:shell_obs}). They are generally perpendicular to the
direction to the center of the SNR and are presumably related to the
shell. This is the area known in the literature as Vela~Y
(\cite{dwa91}; \cite{mil68}).  Directly to the east of Vela~X is the radio
peak known as Vela~Z. This area is confused by the elliptical sidelobe
from the bright \hii\ region RCW~38 (\cite{rod60}), which is not
included in the survey. The area around RCW~38 is included in the
First Epoch Molonglo Galactic Plane Survey (\cite{gre98}). On the
southern side of the central nebula, another region of shell-like
emission (08\h32\h$ -49^\circ00'$) is probably also part of the Vela
SNR.  This coincides approximately with the southern boundary of the
8\arcdeg\ remnant (\cite{asc92}; \cite{dun96}).

The survey contains many unresolved and barely resolved sources, most
of which are presumably background sources.\footnote{About 39
  small-diameter sources above 5~mJy at 843~MHz are found per square
  degree away from the Galactic Plane (\cite{lar94}).} However, some
may be density enhancements in the Vela emission or other Galactic objects, such as compact \hii\ regions, planetary
nebulae, pulsars or small-diameter SNRs. Several of these sources have
unusual coincidences with extended structures. From the survey it is
unclear whether they are in fact background sources, or whether they
are `knots' in the SNR emission.  Follow-up VLA observations
(\cite{boc98a}) of four of these sources have not found evidence for a
Galactic origin.  

The unrelated supernova remnant Puppis A is also contained within the
survey. It is an oxygen-rich SNR of age about 3700~yr (\cite{win88})
at an accepted distance of 2~kpc (\cite{are90}). Puppis A has
previously been imaged separately with the MOST (\cite{kes87};
\cite{are90}). It falls approximately on the 8\arcdeg\ X-ray boundary
of the Vela SNR. However, no Vela radio emission is obvious in the
vicinity.

\subsection{The northern shell}
\label{sec:shell_obs}

The present survey of the Vela SNR covers much of the brightest region
of the Vela shell. The image in figure~\ref{fig:shell_most} is part of
the survey showing the northern section of the Vela SNR shell.  The
following discussion focuses on this area, where the radio emission
from the shell is most prominent and not confused by emission from
unrelated Galactic objects or by artefacts.

The extended structure in this image is in a series of filamentary
arcs across the image at position angles ranging from $70^\circ$ to
170\arcdeg.  The structure is generally concave towards the center of
the SNR: there are no significant radial filaments.  The majority of
the filaments are resolved by the MOST, with widths ranging from
1\arcmin\ to 6\arcmin\ and peak surface brightnesses up to 20~m\jb.
These filaments generally have a sharp edge on the side away from the
center of the remnant, while towards the remnant center they may fade
over several arcminutes. The sharper outside profile is 
consistent with the `projected sheet' picture of filamentary emission
(\cite{hes87}). The effect may also indicate that the filaments
are in fact edges, spatially filtered by the MOST so that only
the sharp transitions appear.

\subsubsection{A multi-wavelength comparison}

The availability of three datasets of comparable resolution at widely
spaced wavelengths gives us an opportunity to understand the spatial
relationship between the underlying physical processes. In addition to
the 843~MHz survey with the MOST, \hal\ and soft X-ray data are
available. The radio image shows primarily the non-thermal synchrotron
emission, the optical filaments are line emission resulting from
recombination in cooling processes, while the X-rays are shock heated
thermal radiation.

An \hal\ image of the northern Vela shell is shown in
figure~\ref{fig:shell_ol}(a), overlaid with a contour (at
approximately the $3\sigma$ level) from the radio image of
figure~\ref{fig:shell_most}.  This image is from a test observation
(kindly made and reduced by M.~S. Bessell) for the MSSSO Wide Field
CCD \hal\ Imaging Survey (Buxton, Bessell, \& Watson
1998\nocite{bux98}).  
The observation was made using a
$2048\times2048$ 24~\micron\ pixel CCD through a 400~mm, f/4.5
Nikkor-Q lens, at the 16-inch (0.4 m) telescope facility at Siding
Springs Observatory. Pixel spacing in the image is 12\arcsec, giving a
field size of 7\arcdeg\ square. The portion of this image presented
here is taken from the central ($5^\circ\times5^\circ$) region, where
vignetting in the 1.5~nm filter is not significant.  The image has
been derived from two frames with a total exposure time of 1400~s,
which were bias-subtracted and flat-field corrected before averaging.
No correction has been made for the effect of cosmic rays.  No
continuum observation was made for subtraction. Consequently, the
image presented here contains stars and a continuum component in the
extended emission.  A coordinate system was applied to the image by
comparison with a Digital Sky Survey image (\cite{mor95}) using the
\texttt{KARMA} package (\cite{goo96}). The registration is within the
resolution of the radio data.

The Vela SNR was observed as part of the ROSAT All-sky Survey between
1990 October and 1991 January. An image of the Vela SNR (0.1--2.4~keV,
with angular resolution 1\arcmin) from the survey has been presented
by Aschenbach et al.\ (1995\nocite{asc95}), and part is reproduced in
figure~\ref{fig:shell_ol}(b), overlaid with the same radio contour as
in figure~\ref{fig:shell_ol}(a).  In the figure, the top section of
the X-ray image (black) is the Galactic background. The surface
brightness at the edge of the SNR shell is
$7\times10^{-15}~\rm{erg~cm^{-2}~s^{-1}~arcmin^{-2}}$ (\cite{asc95}). To
the south, the surface brightness increases by a factor of 500 to the
brightest part (white), which is the most intense X-ray emission region
in the entire SNR. The first grey area ($\delta=-41^\circ40'$) marks
the edge of the main shock, seen in projection.

\subsubsection{Morphological analysis}
\label{sec:shell_morph}

By considering only the radio and \hal\ images
(figure~\ref{fig:shell_ol}(a)), it is possible to see immediately the
most striking aspect of the comparison, namely the contrast between
the optical and radio emission regions. As will be discussed below,
this has a simple theoretical basis, but is contrary to the picture
seen in other SNRs in those cases where optical emission has been
compared with well-resolved radio shell structure.  The brightest
radio filaments are (as noted earlier) generally oriented
perpendicularly to the direction to the SNR center and are without
optical counterparts.  Likewise, many of the optical filaments are
without radio counterparts. However, one of the brightest optical
filaments (with orientation similar to the radio structures), centered
on 08\h36\m$-$42\arcdeg50\arcmin, does have a faint radio counterpart.
By contrast, the equally bright optical filaments in the south-west
corner of the image are without radio counterparts in the MOST image.
These filaments are generally not oriented perpendicularly to the
direction to the SNR center in the same way as the radio filaments.

In addition to the optical filamentary structure, diffuse optical
emission is also present. This is concentrated to the eastern side of
the image, in the general area of the strong radio filaments, but
there is no obvious correlation between the diffuse optical emission
and the radio filaments. No direct measure of the effect of
extinction on the  \hal\ image is available.

The complete X-ray image (\cite{asc95}) shows by
its near-circular shape that it delineates the projected edge of those
parts of the main shock that are still expanding into a relatively
homogeneous medium.  The regions of optical and radio emission
described so far are interior to this main X-ray shell.  At the
western side of the main X-ray boundary in figure~\ref{fig:shell_ol}(b),
we see significant optical and radio emission clearly present
close to the X-ray edge. Here the radio and optical emission agree
quite well, in an arc with apex at 08\h35\m$-$42\arcdeg10\arcmin,
just behind the outer edge of the X-ray
emission. 

The X-ray emission is quite different in form to the emission we see
in the optical and radio regimes. Apart from the main edge, it is
relatively diffuse and smooth. By contrast, the radio and optical
images are dominated by filamentary structure. However, we note that
the radio image has reduced sensitivity to smooth structure, due to
the MOST's spatial response.

The bright optical filament at 08\h36\m$-$42\arcdeg50\arcmin\ traces
the exterior (with respect to the remnant expansion) of the brightest
peaks of the X-ray emission. Yet not all the optical filaments exhibit
this relationship.  The diffuse optical component has no obvious
 X-ray counterpart and is strongest where
the X-ray emission is not quite so bright, to the east.  The radio
filaments are also partially correlated with the X-ray emission.
Several follow changes in X-ray brightness. However, the most central
filament (08\h39\m$-43^\circ10'$) is less well correlated: it crosses
a bright region of X-ray emission.

\subsubsection{Radiation mechanisms}

In SNRs, optical and X-ray emission are both typically due to thermal
processes. However, quite different physical conditions are involved.
Thermal X-ray emission is the result of fast shocks propagating
through a rarefied medium, with density 0.1--1~cm$^{-3}$, shocked to
temperatures of 10$^6$--10$^7$~K (\cite{loz92}). The optical emission
typically observed is produced by hydrogen recombination of cooling
shocked gas at about $10^4$~K, with density a few times
$10^2$~cm$^{-3}$.

One model which has had success explaining optical and X-ray
observations of the Cygnus Loop (\cite{hes86}; \cite{gra95a};
\cite{lev96}) invokes large ($\gtrsim 10^{14}$~m) molecular clouds
with which the expanding shock is interacting. The optical emission
comes from the shocked cloud, where the dense material is not heated
to temperatures as high as those which are maintained in the less
dense X-ray emitting regions.  This emission is due to recombinative
cooling after the passage of the shock. Behind the optical emission,
the X-ray emission is further brightened by the passage of a reflected
(or reverse) shock due to the density contrast between the cloud and
the less dense inter-cloud material. Where the main shock does not
encounter molecular clouds, we do not expect to see recombinative
cooling. Instead the non-radiative shock may be traced by fainter
Balmer filaments (Hester, Raymond, \& Blair 1994\nocite{hes94}).

The present observations of the northern Vela shell fit nicely into
this picture.  If the majority of optical emission was from the main
shock interacting with a relatively uniform medium, but seen here in
projection, we would expect also to see it along the entire edge of
the X-ray shell, where accentuation of sheet-like emission in
projection would be strongest. This is not observed, implying the
emission is localized and due to some interaction in density inhomogeneities
with a filling factor much less than unity. The cloud interaction
model  is further supported by the
presence of X-ray brightened regions (figure~\ref{fig:shell_ol}(b))
immediately behind the bright optical filamentary structure centered
on 08\h36\m$-$42\arcdeg50\arcmin\ in
figure~\ref{fig:shell_ol}(a).  We might be seeing this emission in
projection, significantly in front of or behind the plane of the
explosion center transverse to the line of sight. This would indicate
local density enhancements very close to the main shock.
Alternatively, it could be nearly in the plane of the explosion
center, with a shock velocity significantly reduced by interactions
with more dense material. Some of the emission could be from regions
already passed and energized by the main shock.

The digital 60~\micron\ images in the IRAS Sky Survey Atlas
(\cite{whe94}) support the thermal emission model for the X-ray
emission. Much of the X-ray structure does have an infra-red
counterpart. However, infra-red images are generally less useful
thermal diagnostics than X-ray images near the Galactic Plane, since
the infra-red observations are dominated by diffuse Galactic emission
and confusion from other sources (\cite{whi91}).

An alternative model for SNR optical/X-ray emission (\cite{mck75}),
explains SNRs with centrally-peaked X-ray emission (\cite{whi91}). In
this model cold dense clouds with a small filling factor have been
passed by the main shock and are evaporating by conductive heating
from the postshock gas.

Both these models rely on molecular clouds to explain the observed
features.  Molecular clouds have been detected in the direction of the
Vela SNR (May, Murphy, \& Thaddeus 1988\nocite{may88}). The initial
survey was of $^{12}$CO and $^{13}$CO $J=1\rightarrow0$ line emission
with a resolution of 0\fdg5.  Higher resolution follow-up observations
(\cite{mur91}) covered only the eastern part of the Vela SNR shell. A
cloud with a barely resolved peak at 08\h41\m$-41^\circ20'$ is seen,
with a distance estimated to be 0.5--2.0~kpc, i.e.\ immediately behind
the Vela SNR. However, this cloud appears coincident with a bright
\hii\ region seen optically to the north of
figure~\ref{fig:shell_ol}(a), and might not be responsible for the
observed optical features in the Vela shell. \hi\ may be a better
tracer of density in the Vela shell region. Dubner et al.\ 
(1998\nocite{dub98}) find a near-circular shell of \hi\ surrounding
the northern edge of the remnant, with column densities up to
$10^{21}$~cm$^{-2}$, and estimate the pre-shock gas to have had a
density of 1--2~cm$^{-3}$. The \hi\ shell traces the X-ray edge of the
remnant, enclosing the radio and optical filaments.
 
In the simple radio emission model for the interaction of supernova
explosions with the ISM (\cite{wol72}), Vela is in the
radiative or snowplow phase of evolution, having swept
up significant matter and dissipated much of the original kinetic energy of
the explosion. A cool dense shell surrounds a hot interior. This model can
account for the faint radio emission seen just behind the X-ray edge,
which indicates the presence of compressed magnetic fields and
accelerated particles, probably from the diffusive shock mechanism
(\cite{ful90}). It does not account for the brighter localized
filaments apparently well behind the main shock.

Duin \& Van Der Laan (1975\nocite{dui75}) present a consistent picture
for the coincidence of radio and optical emission which is observed in
``middle-aged'' shell remnants. This model, based on observations of
IC443, proposes that the magnetic field required for synchrotron
emission is frozen into condensations forming in the cooling
instabilities which then give rise to the optical emission. We do not
find significant radio/optical coincidence in our Vela observations.
Consequently, if this process is occurring, then we can infer that the
cooling material around the radio filaments is not at an appropriate
temperature for the emission of recombination radiation. One
explanation is for a long period to have elapsed following the passage
of the radiative shock, allowing substantial cooling while still
preserving the conditions for synchrotron emission (\cite{bla82}).
Alternatively, where shock-accelerated particles producing the optical
filaments are located, the magnetic field may not be sufficiently
compressed to cause detectable synchrotron emission.

The applicability of these models may be investigated further with
magnetic field information, provided by polarimetry.  Polarized
intensity has been observed in the Vela shell (\cite{dun96}), but high
resolution measurements at several frequencies will be required to
examine the magnetic field structure in this area in detail. Blandford
and Cowie (1982\nocite{bla82}) note that individual filaments ought to
be polarized parallel to their longest dimensions, although they may
be too faint to be detected with current instruments.

Good agreement between optical and radio emission has been found in
other middle-aged shell SNRs such as IC443 (\cite{dui75}), the Cygnus
Loop (\cite{str86}) and HB3 (\cite{fes95}).  The situation in Vela is
quite different and the reason for this is not apparent.  Extinction
may be a culprit, obscuring some of the \hal\ emission.  However, the
coincidence of diffuse optical emission with bright radio filaments,
noted above, argues against massive extinction in this direction.

This initial investigation of the optical/radio/X-ray correlations in
the region indicates that a fuller investigation would be profitable.
A first step would be to obtain optical spectral information to
separate non-radiative and radiative filaments, allowing a detailed
comparison with the model of Hester \& Cox (1986\nocite{hes86}).

\subsection{Vela~X}
\label{sec:most_x}

A view of the central nebula of the Vela SNR is shown as a greyscale image
in figure~\ref{fig:velax.843} and as a ruled surface plot in
figure~\ref{fig:velax_hid}.  Each representation emphasizes different
characteristics. The greyscale image gives a good overall view of the
region, while the ruled surface plot helps to show the nature of the
filamentary structure and highlights the small-diameter sources.

The first thing to note in the images is that at the resolution of
these observations the nebula is seen to be composed of many filaments
or wisps, at a variety of orientations and on many angular
scales. Several of the brighter filaments are aligned approximately
north-south. It is important to realize that the flux density detected in
this image is
only a small fraction of the total flux density of the remnant,
because of absent low spatial frequency information. The total flux density
of the extended features in the MOST image of Vela~X is calculated to be
$28\pm2$ Jy, which becomes 130 Jy when correction is made for the negative
bowl artefact surrounding the nebula. This is approximately 12\% of the
estimated single dish flux density of Vela~X (\cite{dwa91}). One benefit of the
MOST acting as a spatial filter is the prominence it gives to smaller scale
structures with size of the order of the X-ray feature seen
by Markwardt and \"Ogelman (1995). In figure~\ref{fig:velax.843}, the
central radio filament overlaid on the X-ray feature by Frail et al.\ 
(1997) is marked `1'. This filament does not look strikingly different
from other filaments in the region, for example the filament marked
`2'. However, we see in the 8~GHz Parkes image of Milne (1995) that
filament `1' is located at the brightest part of the Vela~X nebula. Frail
et al.\ (1997) have argued that this radio filament may be associated with the
X-ray feature, but it is morphologically indistinguishable from other
filaments in the image.  The central radio filament looks so prominent
in the 327~MHz VLA image of Frail el al.\ (1997) partially because
that image is uncorrected for the VLA primary beam attenuation at the
edge of the field. Also, the maximum entropy method of deconvolution used
for the VLA data promotes the flux density at low
spatial frequencies more than the CLEAN algorithm used to deconvolve the
MOST observations.

Several further interesting objects in the region should be noted. In
figure~\ref{fig:velax.843} a filament (`3') extends through the pulsar
position (at the head of the arrow) to the south and may connect to
filament `1'. The axis of symmetry of these two filaments is closely
aligned with the direction of motion of the pulsar (\cite{bai89}),
shown on the image with an arrow.  Using the proper motion measurement
of Bailes et al., we notice that over its lifetime (assuming an age of
12,000~yr) the pulsar has moved to its present position from a bright
region to the south-east. An excess of high-energy $\gamma$-rays has
been detected from near this putative birthplace (\cite{yos97}).
Greater age estimates (e.g.\ \cite{asc95}; \cite{lyn96}) change this
slightly as they increase the distance moved by the pulsar by up to a
factor of two. Just to the north of the pulsar is a 3\arcmin\ 
crescent-shaped synchrotron nebula, seen originally by Bietenholz,
Frail, \& Hankins (1991)\nocite{bie91}.  They resolved out the
extended structure in the region, whereas here we see that the
crescent is one bright region of much extended emission around the
pulsar.  Some very faint structures found around the edge of Vela~X
appear unusual. Object `4' has a shape reminiscent of many shell
supernova remnants, but if it is associated with Vela~X it might be a
blowout from the nebula. Object `5' is a faint streamer apparently
connecting Vela~X to the Vela shell (cf. figure~\ref{fig:survey}). It
might be argued that this is actually a foreground or background
projection of the surface of a shock `bubble', but it is substantially
thinner than any of the shell filaments.
  
\section{Conclusion}

The radio survey presented in this paper contains the highest
resolution observations yet made of the bulk of the Vela supernova
remnant and resolves the structure of the remnant in more detail than
has been possible for any other composite remnant. The resolution of
this observation of the Vela~X region is a factor of two greater than
that presented by Frail et al.\ (1997\nocite{fra97}), and covers the
entire plerion, unaffected by primary beam attenuation. The Vela
plerion in the radio consists both of diffuse and filamentary
emission. Although the survey does not contain information on the
largest spatial scales, this structure may be inferred from single
dish observations at higher frequencies (\cite{mil95}; \cite{dun96}),
which show that the filamentary emission in the survey covers the same
area as the more diffuse emission from Vela~X seen in total power
images.  The region immediately surrounding the Vela pulsar contains
much non-thermal emission in addition to the possible pulsar wind
nebula seen by Bietenholz et al.\ (1991).

The two distinct regions of the Vela SNR, the shell and the plerion,
have in the past been considered separate entities because of their
different spectral indices. This characteristic puts Vela in the
composite class with SNRs such as G$326.3-1.8$ (MSH~$15-56$: Clark,
Green, \& Caswell 1975\nocite{cla75}; \cite{whi96}) and G$0.9+0.1$
(\cite{hel87}). In the images presented in this paper, we now see the
shell and plerion in fine detail and they separately show strong
similarities with what we see in other SNRs, observed at similar
resolutions. The shell filaments are comparable to those seen in the
Cygnus Loop (\cite{gre84}; \cite{str86}), oriented perpendicularly to
the direction to the SNR center, with \hal\ and X-ray counterparts. By
contrast, the filamentary structure within the plerion is more
nebulous and has a gross alignment approximately north-south. Its
appearance is reminiscent of the filaments in the Crab Nebula
(\cite{bie90}).  Thus we can identify both a shell and a plerion
within the Vela SNR, classifying it unambiguously as a composite
remnant.

We propose that it be considered as the archetypal Galactic member of
the composite class.  In angular extent, the Vela SNR is respectively
50 and 12 times larger than G$0.9+0.1$ and G$326.3-1.8$, allowing detailed
studies at a variety of wavelengths. It
is now appropriate to use this object as a key laboratory for
studying the properties of SNRs and the interstellar medium.

The investigation of the shell of the Vela SNR in this paper focussed
on its northern side. Like parts of the Cygnus Loop, this region can
be explained by a model of a fast shock heating interstellar material
to X-ray emitting temperatures and interacting with denser clouds to
produce \hal\ recombination line emission (\cite{hes86};
\cite{gra95a}; \cite{lev96}).  The expanding shock also produces
bright non-thermal radio emission not well correlated with these \hal\ 
filaments, in contrast with the optical/radio agreement seen in many
other middle-aged SNRs.  To continue investigating this region,
observations of other optical emission lines are needed to separate
projected Balmer filaments, produced at the outer shock, from
recombination line emission at molecular cloud interactions.
High-resolution polarization observations of the radio shell filaments
are the obvious next step to investigate the magnetic field associated
with the non-thermal emission.

Further study of the remnant's plerionic component, Vela~X, should
determine how the Vela pulsar transfers its rotational kinetic energy
to the nebula. With an age at least ten times that of the Crab Nebula,
we might expect the Vela plerion to show evolutionary trends in the
relative emission strengths in different wavelength regimes. The
absence of obvious correlations between radio emission and the optical
filaments (Elliott, Goudis, \& Meaburn 1976\nocite{ell76}) already
contrasts Vela~X with the Crab Nebula, where the radio filaments
surround the optical filaments (\cite{bie91a}). The recent discovery
of a possible X-ray `jet', which might be the conduit for energy
transfer to the nebula from the pulsar (\cite{mar95}), further
contrasts Vela~X with other plerions currently known.

\acknowledgements

The Molonglo Observatory Synthesis Telescope is operated by the School
of Physics, with funds from the Australian Research Council and the
Science Foundation for Physics within the University of Sydney.  The
authors thank D. A. Frail for useful discussions in the course of this
work; J. E. Reynolds for assistance with the pulsar gating
observations; B. Aschenbach and M. S. Bessell for providing
electronic versions of their images; and M. Bailes, P. M. McCulloch and
R. N. Manchester for providing Vela pulsar timing data.
D. C.-J. B. also acknowledges financial support from an
Australian Postgraduate Award while at the University of Sydney.

\nocite{dwa91,mil68,asc95,bux98} 

\clearpage

\figcaption{The 843~MHz survey of the Vela SNR. Vela~X is the
  central, nebulous region composed of many filaments in an area about
  2\arcdeg\ across, centered on 08\h35\m $-45$\arcdeg30\arcmin, with
  the Vela pulsar (suppressed in this image) offset from the center of
  the nebula at a position marked with a cross.  The radio shell can
  be seen most clearly around 08\h42\m $-43^\circ00'$. The separate
  SNR, Puppis A, is also included in the survey, at
  08\h$22^m-43^\circ00'$. The image greyscale is saturated outside the
  intensity range of $-20$ to 30 m\jb\ to show the extended structure
  most clearly.
  \label{fig:survey}}

\figcaption{A cartoon of the same area as figure~\ref{fig:survey},
  showing key features in the region. The radio peaks (in total power
  observations) of Vela~X, Y and Z are marked at the positions adopted
  in the literature (Dwarakanath 1991; Milne 1968a). The names Vela~Y
  and Z are superseded now that these regions have been resolved to
  show the filamentary structure. The approximate boundary of the
  X-ray remnant (Aschenbach et al.\ 1995) is marked with a dotted line.
  \label{fig:vela_cartoon}}

\figcaption{A sub-image of figure 1, showing the northern
  radio continuum (843~MHz) filaments.  The intensity scale is linear,
  from $-10$ (white) to 15 (black) m\jb.
  \label{fig:shell_most}}

\figcaption{A single 843~MHz contour (at 4~m\jb) on (a) \hal\ and (b)
  X-ray images. The \hal\ image is from the MSSSO \hal\ Survey (Buxton
  et al.\ 1998), courtesy of M.\ S.~Bessell. The X-ray image was kindly
  provided by B. Aschenbach (originally published in Aschenbach et
  al.\ 1995). 
  \label{fig:shell_ol}}

\figcaption{A sub-image of figure 1, showing Vela X, the central nebula
  of the Vela supernova remnant.  The arrow shows the direction and
  magnitude of the proper motion of the pulsar over 12,000~yr,
  assuming constant velocity as measured by Bailes et al.\ (1989). The
  arrow's head is at the present location of the pulsar (suppressed in
  this image). The numbers refer to features discussed in the text.
  \label{fig:velax.843}}

\figcaption{A ruled surface plot of the same area as
  figure~\ref{fig:velax.843}, emphasizing the disordered nebulosity of
  Vela~X and highlighting the relative strengths of the point sources
  and the extended structure. The data have been convolved to a
  circular beam of $120''\times 120''$ to reduce the noise, which also
  accentuates the extended emission (by a factor of 5.6) relative to
  the compact sources.
  \label{fig:velax_hid}}

\hspace*{1cm}

\noindent An alternative version of this preprint, which contains
higher-quality postscript figures, is available at
\texttt{http://astro.berkeley.edu/\~{ }dbock/papers/} .

\end{document}